\input harvmac
%\draftmode
\noblackbox

\def\p{\partial}

%% MACROS
\font\cmss=cmss10
\font\cmsss=cmss10 at 7pt
\def\IL{\relax{\rm I\kern-.18em L}}
\def\IH{\relax{\rm I\kern-.18em H}}
\def\IR{\relax{\rm I\kern-.18em R}}
\def\inbar{\vrule height1.5ex width.4pt depth0pt}
\def\IC{\relax\hbox{$\inbar\kern-.3em{\rm C}$}}
\def\rlx{\relax\leavevmode}

\def\ZZ{\rlx\leavevmode\ifmmode\mathchoice{\hbox{\cmss Z\kern-.4em Z}}
 {\hbox{\cmss Z\kern-.4em Z}}{\lower.9pt\hbox{\cmsss Z\kern-.36em Z}}
 {\lower1.2pt\hbox{\cmsss Z\kern-.36em Z}}\else{\cmss Z\kern-.4em
 Z}\fi} 
%%% misc.
\def\IZ{\relax\ifmmode\mathchoice
{\hbox{\cmss Z\kern-.4em Z}}{\hbox{\cmss Z\kern-.4em Z}}
{\lower.9pt\hbox{\cmsss Z\kern-.4em Z}}
{\lower1.2pt\hbox{\cmsss Z\kern-.4em Z}}\else{\cmss Z\kern-.4em
Z}\fi}

\def\CN {{\cal N}}

\def\CL {{\cal L}}

\def\CO {{\cal O}}

\def\CS {{\cal S}}

%% MORE MACROS

\def\CN {{\cal N}}

\def\CO {{\cal O}}

\def\CS {{\cal S }}

\def\Tr{{\rm Tr}}

\def\zb {\bar{z}}

\def\qb {\bar{q}}
\font\manual=manfnt 
\def\dbend{\lower3.5pt\hbox{\manual\char127}}

\def\IZ{\relax\ifmmode\mathchoice
{\hbox{\cmss Z\kern-.4em Z}}{\hbox{\cmss Z\kern-.4em Z}}
{\lower.9pt\hbox{\cmsss Z\kern-.4em Z}} {\lower1.2pt\hbox{\cmsss
Z\kern-.4em Z}}\else{\cmss Z\kern-.4em Z}\fi}
\def\half {{1\over 2}}
\def\sdtimes{\mathbin{\hbox{\hskip2pt\vrule height 4.1pt depth -.3pt
width .25pt \hskip-2pt$\times$}}  }

%bat. Bye.
%Ram

\def\p{\partial}
\def\pb{\bar{\partial}}

\def\bar{\overline}
\def\CS{{\cal S}}
\def\CN{{\cal N}}

\def\rt2{\sqrt{2}}
\def\irt2{{1\over\sqrt{2}}}
\def\rtt{\sqrt{3}}

\def\t{\tilde}
\def\ndt{\noindent}

\lref\kutsei{D.~Kutasov and N.~Seiberg, ``Non-critical superstrings'', Phys.\ Lett.\ B {\bf 251}, 67 (1990).}

\lref\kutgiv{A.~Giveon and D.~Kutasov, ``Little string theory in a double scaling limit'', JHEP {\bf 9910} 34 (1999) [arXiv:hep-th/9909110]. }

\lref\kutgivads{A.~Giveon and D.~Kutasov, ``Notes on $AdS_3$'', Nucl.\ Phys.\ B {\bf 621}, 303 (2002)  [arXiv:hep-th/9909110]. }

\lref\kutgivpelc{A.~Giveon, D.~Kutasov and O.~Pelc, ``Holography for noncritical superstrings'', JHEP {\bf 9910} 35 (1999) [arXiv:hep-th/9907178]. }

\lref\kutasov{D.~Kutasov, ``Some properties of (non)critical strings'', [arXiv:hep-th/9110041].}

\lref\wittenbh{E.~Witten, ``String theory and black holes'', Phys.\ Rev.\ D {\bf 44}, 314 (1991) \semi 
G.~Mandal, A.~M.~Sengupta and S.~R.~Wadia, ``Classical solutions of two-dimensional string theory'', Mod.\ Phys.\ Lett. A {\bf 6}, 1685 (1991) \semi
S.~Elitzur, A.~Forge and E.~Rabinovici, ``Some global aspects of string compactifications'', Mod.\ Phys.\ Lett. A {\bf 6}, 1685 (1991).}

\lref\dvv{R.~Dijkgraaf, H.~Verlinde and E.~Verlinde, ``String propagation in a black hole geometry'',  Nucl.\ Phys.\ B {\bf 371}, 269 (1992).}

\lref\eguchi{T.~Eguchi and Y.Sugawara, ``Modular invariance in superstring on Calabi-Yau n-fold with A-D-E singularity'',  Nucl.\ Phys.\ B {\bf 577}, 3 (2000) [arXiv:hep-th/0002100].  }

\lref\mizo{S.~Mizoguchi, ``Modular invariant critical superstrings on four-dimensional Minkowski space $\times$ two-dimensional black hole'', JHEP {\bf 0004} 14 (2000) [arXiv:hep-th/0003053]. }

\lref\bilal{A.~Bilal and J.~L.~Gervais, ``New critical dimensions for string theories'', Nucl.\ Phys.\ B {\bf 284}, 397 (1987). }

\lref\farkas{H.~M.~Farkas and I.~Kra, ``Theta constants, Riemann surfaces and the modular group'', Graduate studies in mathematics, Vol.37. Amer.\ Math.\ Soc.\ }

\lref\horikap{K.~Hori and A.~Kapustin, ``Duality of the fermionic 2-d black hole and $\CN=2$ Liouville theory as mirror symmetry'', JHEP {\bf 0108} 045 (2001) [arXiv:hep-th/0104202].}

\lref\oogvafa{H.~Ooguri and C.~Vafa, ``Two-dimensional black hole and singularities of CY manifolds'', Nucl.\ Phys.\ B {\bf 463}, 55 (1996) [arXiv:hep-th/9511164]. }

\lref\dilholo{O.~Aharony, M.~Berkooz, D.~Kutasov and N.~Seiberg, ``Linear dilatons, NS5-branes and holography'', JHEP {\bf 9810} 004 (1998) [arXiv:hep-th/9808149]. }

\lref\seiberg{N.~Seiberg, ``Notes on quantum Liouville theory and quantum gravity'', Prog.\ Theor.\ Phys.\ Suppl {\bf 102} 319 (1990).}

\lref\maldoog{J.~Maldacena and H.~Ooguri,``Strings in $AdS_3$ and $SL_2(\IR)$ WZW model 1: The spectrum'', J.\ Math.\ Phys.\ {\bf 42} 2929 (2001)  [arXiv:hep-th/0001053]. }

\lref\fms{D.~Friedan, E.~Martinec and S.~Shenker, ``Conformal invariance, supersymmetry and string theory'', Nucl.\ Phys.\ B {\bf 271}, 93 (1986). }

\lref\chs{C.~Callan, J.~Harvey and A.~Strominger, ``Supersymmetric string solitons'', [arXiv:hep-th/9112030], in Trieste 1991, Proceedings, String theory and Quantum Gravity 1991, 208.}

\lref\polchinski{J.~Polchinski, ``String theory, Vol. I, II'', Cambridge University Press (1998).}

\lref\kleb{C.~P.~Herzog, I.~R.~Klebanov and P.~Ouyang, ``Remarks on the warped deformed conifold'',  [arXiv:hep-th/0108101].}

\lref\seiwitt{N.~Seiberg and E.~Witten, ``The $D1/D5$ system and singular CFT'', JHEP {\bf 9904} 017 (1999), [arxiv:hepth/9903224].}

\lref\mukhi{K.~Dasgupta and S.~Mukhi,``Brane constructions, conifolds and M-theory'', Nucl.\ Phys.\ B {\bf 551}, 204 (1999), [arxiv:hepth/9811139]. }

\lref\tong{D.~Tong, ``NS5-branes, T-duality and worldsheet instantons'', JHEP {\bf 0207} 013 (2002) [arXiv:hep-th/0204186].}

\lref\juan{J.~Maldacena, Private communication.}

\lref\fzz{V.~A.~Fateev, A.~B.~Zamolodchikov and Al.~B.~Zamolodchikov, Unpublished notes.}

\lref\kazsuz{Y.~Kazama and H.~Suzuki, ``New $N=2$ superconformal field theories and superstring compactification'', Nucl.\ Phys.\ B {\bf 321}, 232 (1989). }

\lref\harmoo{R.~Gregory, J.~Harvey and G.~Moore,``Unwinding strings and T-duality of Kaluza-Klein and H-monopoles'', Adv.\ Theor.\ Math.\ Phys.\ {\bf 1} 283 (1997), [arXiv:hep-th/9708086]. }

\lref\klemvafa{A.~Klemm, W.~Lerche, P.~Mayr, C.~Vafa and N.~Warner, ``Self-dual strings and $N=2$ supersymmetric field theory'', Nucl.\ Phys.\ B {\bf 477}, 746 (1996),  [arXiv:hep-th/9604034]\semi
S.~Gukov, C.~Vafa and E.~Witten, ``CFT's from Calabi-Yau four-folds'', Nucl.\ Phys.\ B {\bf 584}, 69 (2000),  [arXiv:hep-th/9906070]. }

\lref\polya{A.~M.~Polyakov, ``Quantum Geometry of Bosonic Strings'', Phys.\ Lett. B {\bf 103}, 207 (1981).}

\Title{\vbox{\baselineskip12pt \hbox{hep-th/0305197}\hbox{PUPT-2086}
}} {\vbox{ {\centerline {Notes on Non-Critical Superstrings}} 
{\centerline { in Various Dimensions  }} 
}}

\centerline{Sameer Murthy\footnote{$^1$}{smurthy@princeton.edu}
}
\smallskip
\centerline{\sl Department of Physics,  Princeton University}
\centerline{\sl Princeton, NJ 08544, USA}
\medskip

\vskip .3in
\centerline{\bf Abstract}
We study non-critical superstrings propagating in $d \le 6$
dimensional Minkowski space or equivalently, superstrings propagating on the two-dimensional Euclidean black hole tensored with $d$-dimensional Minkowski space. We point out a subtlety in the construction of supersymmetric theories in these backgrounds, and explain how this does not allow a consistent geometric interpretation in terms of fields propagating on a cigar-like spacetime. We explain the global symmetries of the various theories by using their description as the near horizon geometry of wrapped NS5-brane configurations. In the six-dimensional theory, we present a CFT description of the four dimensional moduli space and the global $O(3)$ symmetry. The worldsheet action invariant under this symmetry contains both the $\CN=2$ sine-Liouville interaction and the cigar metric, thereby providing an example where the two interactions are naturally present in the same worldsheet lagrangian already at the non-dynamical level. 

\Date{May, 2003}

\newsec{Introduction and summary}

It has been known for a long time that consistent string theories can live in low number of dimensions. These theories typically develop a dynamical Liouville mode  \polya\ on the worldsheet, and have been called non-critical string theories. These theories can be thought of as Weyl invariant string theories in a spacetime background which is one dimension higher, and contains a varying dilaton and a non-trivial `tachyon' profile which corresponds to the Liouville interaction on the worldsheet. While the bosonic non-critical string is perturbatively consistent for one or less spacetime dimensions, the authors of \kutsei\ constructed non-critical theories with $\CN=(2,2)$ supersymmetry on the worldsheet, which have the $\CN=2$ super-Liouville, also known as the sine-Liouville interaction, on the worldsheet giving rise to consistent string theories with spacetime supersymmetry in all even dimensions less than ten. 

One aspect of this construction which has not been completely clear is the geometric interpretation of these theories. Fateev, Zamolodchikov and Zamolodchikov \fzz\ conjectured that the sine-Liouville theory is equivalent to the conformal field theory describing the two dimensional Euclidean black hole or the cigar \wittenbh . This conjecture was extended to the $\CN=2$ supersymmetric case by \kutgiv\ and proved using the techniques of mirror symmetry by \horikap . String theory in the black hole background has been previously studied by many authors [see e.g.\dvv]. This duality provides a possible interpretation of the non-critical superstrings in $d$ spacetime dimensions as an (infinite) set of fields propagating on the cigar tensored with $\IR^{d-1,1}$, analogous to string theory in ten flat dimensions. 

We shall study the non-critical theories in flat space for various values of $d$ using the above two dual worldsheet conformal field theories:
\item{1.} The $\CN=2$ supersymmetric sine-Liouville theory tensored with flat spacetime\foot{$d=0,2,4,6$; $d=0$ is interpreted as the pure sine-Liouville theory. For $d=8$, the extra two dimensions are flat, producing ten dimensional flat space string theory.} $\IR^{d-1,1}$ as defined by \kutsei .
\item{2.} The  $\CN=2$ supersymmetric version of the cigar defined as the Kazama-Suzuki \kazsuz\ supercoset $SL_2(\IR)/U(1)$, tensored with $\IR^{d-1,1}$.

More recently, it has been understood how these theories fit into the moduli space of superstring theories. Motivated by the search for a holographic description of these theories, \refs{\kutgiv , \kutgivpelc } conjectured that the non-critical theories arise as a certain double scaling limit of ten dimensional string theory. One approaches a point in moduli space of string theory on a Calabi-Yau manifold where it develops an isolated singularity, taking the string coupling to zero at the same time in such a way as to keep a combination of the two parameters fixed. To study the theory of the singular Calabi-Yau in the limit of $g_s \rightarrow 0$, one replaces the Calabi-Yau by its (non-compact) form near the singularity. To study the double scaling limit, one smoothes out the singularity by deforming the non-compact surface. The precise descriptions of the non-critical theories defined above is:
\item{3.} Superstring theory on $\IR^{d-1,1}$ tensored with the non-compact manifold $\sum_{i=1}^n z_i^2 =\mu$, $n=(12-d)/2$, $z_i \in \IC$. 

\ndt A T-dual \refs{\oogvafa, \klemvafa, \mukhi} of this description is given in terms of wrapped NS5-branes: 
\item{4.} Superstring theory in the near-horizon background of NS5-branes with $d$ flat spacetime directions and $6-d$ directions wrapped on $\sum_{i=3}^n z_i^2 = \mu$.

\ndt The appearance of NS5-branes is not surprising considering that the near horizon geometry of a stack of NS5-branes involves an infinite tube with the dilaton varying linearly along the length of the tube. It has also been noted \harmoo\ that singular geometries like (3) with $\mu=0$\foot{Note that all the four descriptions presented above are singular at $\mu=0$ - the first two descriptions in this limit have an infinite tube with a linearly growing dilaton, and the latter two have geometric singularities at the point $z_i=0$. Turning on $\mu$ resolves the singularities - in (1), the sine-Liouville interaction is turned on, which provides a potential preventing strings from falling into the strong coupling region; in (2), the topology of the tube is changed to that of a cigar with the string coupling at the tip of the cigar determined by $\mu$, thus eliminating the strong coupling singularity; in (3) and (4), the geometric singularities are smoothed out by the deformation. } 
also involve an infinite tube for the winding modes. In both the descrptions (3) and (4), some of the ten dimensions decouple in the limit - roughly speaking, (3) describes the string modes which are localised near the singularity and these only fluctuate in $d+2$ dimensions. 

The purpose of this note is to use the above four descriptions in order to study and clarify some properties of the non-critical theories, in particular their geometric structure. The worlsheet description (1) gives us a handle on perturbative calculations like the spectrum. Using this, we find that there is a clear geometric interpretation in terms of a set of fields propagating on the cigar (2) {\it except} for the appearance of some discrete symmetries which are non-geometric from the cigar viewpoint. On the other hand, we show that these symmetries are natural from the point of view of descriptions (3) and (4). Indeed, they are simply the global symmetries of the above 5-brane (or Calabi-Yau) configurations which do not decouple after taking the limit described above. Our analysis also uncovers some new features of some of the sine-Liouville--cigar duality. 

As just mentioned, in general, the physical spectrum cannot be interpreted as a set of fields propagating on a perhaps singular cigar-like manifold. From the point of view of the worldsheet CFT, this is due to the fact that the chiral $\IZ_2$ symmetry used in the GSO projection is not the naive one of changing the sign of the chiral fermions, rather it acts by translation on the chiral part of the compact boson of the cigar in addition to the above action on the fermions. This is the only consistent choice for the GSO projection because the pure chiral rotation of the fermions on the worldsheet is anamolous due to the curvature of the cigar, and the conserved chiral $U(1)$ current acts on the compact boson of the cigar in addition to the fermions. This means that for a given field on the $d+2$ dimensional geometry, its spin in flat space is correlated with its momentum around the cigar. In the Green-Schwarz picture, this is due to the fact that the conserved current corresponding to the momentum around the cigar is a combination of the naive momentum and a piece acting on the worldsheet fermions. 

The theories we consider asymptote to a linear dilaton geometry and are conjectured \dilholo\ to have holographic non-gravitational duals. In agreement with this statement is the structure of the target space supersymmetry in these theories; the supercharges anticommute to the flat spacetime\foot{In \kutsei , the dilaton direction was interpreted as the time direction, and this was called space-supersymmetry.} momentum generators. We demonstrate that the spectrum can be classified as current multiplets in the boundary theory, as consistent with the holographic interpretation. The conserved $U(1)$ momentum around the cigar is part of the $R$ symmetry in the boundary theory. We shall exhibit this for the first few Kaluza-Klein modes for the $d=4$ theory.

For $d=6$, there is an explicit CFT interpretation as the near-horizon geometry of two parallel non-coincident NS5-branes  \oogvafa . The CFT describing the near-horizon geometry of $k \ge 2$ coincident five branes is $\IR^{5,1}\times \IR \times SU(2)_k$ \chs . There is an infinite throat along which the dilaton increases indefinitely as one approaches the location of the branes, and a trasverse 3-sphere. This theory on $k$ coincident NS5-branes has the global rotation symmetry $SO(5,1) \times SO(4)$. Separating the 5-branes in the four transverse directions in a ring-like structure partially smoothes out the singularity, and the resulting theory has a symmetry $SO(5,1)\times U(1) \times Z_k$, and is conjectured to be string theory on $R^{5,1}\times {SL_2(\IR)_k \over U(1)} \times {SU(2)_k \over U(1)}$.\foot{The GSO projection relates the different factors and so the product is not direct.} \foot{The WZW model corresponding to the sphere consists of bosonic $SU(2)$ currents at level $k-2$ and fermionic currents at level $2$.} In the case $k=2$, the bosonic sphere and the flux disappear and the deformed geometry is precisely the one we want to study for $d=6$. 

This geometric description of the $d=6$ theory reveals some new features of the sine-Liouville--cigar duality. The moduli space of the theory is $\IR^4/\IZ_2$ corresponding to the separation of the two 5-branes, and the global symmetry of this configuration is $SO(4)$ broken to $O(3)$. We shall discuss these features in the conformal field theory. We find that the action of the CFT in the curved directions is not purely the sine-Liouville action, or the cigar action - only specific linear combinations of the two preserves the global symmetry. The $d=6$ theory thus is an example where the duality between the sine-Liouville and the cigar is present already at the kinematic level, giving a better understanding of the duality which was conjectured based on dynamical reasons.

The plan of the paper is the following. In section 2, we shall begin by reviewing the Euclidean black hole background and the non-critical superstring construction. Then we shall lay down the general features of the construction for all the dimensions. In section 3, we describe the special features of all the theories on a case-by-case basis. In particular, we describe many interesting features of the $d=6$ theory. We also make a short note on the $\CN=(4,4)$ algebra in this case. In section 4, we present the explanation of the global symmetries of the various theories using the embedding in ten dimensional flat space string theory. In Appendices A and B, we record the spectrum and one loop partition function of the various theories. Appendix C summarizes the details of the Green-Schwarz formalism, and Appendix D discusses the details of the conformal field theory at second order. 

\newsec{Superstring theories on the Cigar}

The spacetime directions are $X^a = \rho, \theta, X^\mu $, $(\rho\ge 0,\mu=0,1..d-1)$. The geometry in the string frame\foot{We shall set $\alpha' =2$ throughout this paper} is that of a cigar tensored with flat spacetime:
\eqn\metric{\eqalign{
& ds^2= d\rho^2+\tanh^2({Q\rho\over 2}) d\theta^2 +dX^\mu dX_\mu, \qquad \theta \sim \theta +{4\pi\over Q}; \cr
& \Phi = - \log \cosh({Q\rho\over 2}), \qquad B_{ab}=0. \cr
}}
with the string coupling $g_s=e^\Phi$. This metric is a good one for string propagation because the dilaton obeys the equation $2D_aD_b\Phi + R_{ab}=0$, where $D_a$ is the spacetime covariant derivative, and $R_{ab}$ is the spacetime curvature. 

In the asymptotic region $\rho \rightarrow \infty$, the geometry reduces to $\IR^{d-1,1}$ tensored with a cylinder of radius $R={2\over Q}$ with the dilaton varying linearly along its length. When $d+2$ fermions are added to this theory, it also has $\CN=2$ supersymmetry. The currents of the cigar part of the theory in this region are
 \eqn\ntwowss{\eqalign{
 &T_{\rm cig}=-\half (\partial \rho)^2 -\half (\partial\theta)^2 - \half
 (\psi_\rho\partial \psi_\rho + \psi_\theta \partial \psi_\theta)
 - \half Q \partial^2\rho \cr
 &G^\pm_{\rm cig} ={i\over 2} (\psi_\rho \pm i\psi_\theta)\partial(\rho \mp
 i\theta) +{i\over 2} Q\partial (\psi_\rho \pm i\psi_\theta)\cr
 &J_{\rm cig}=-i\psi_\rho\psi_\theta +iQ\partial \theta \equiv i\partial H
 +iQ\partial\theta \equiv i\partial \phi}}
In the exact cigar background the $\CN=2$ supersymmetry is preserved but the exact expressions are more complicated. 
Away from the asymptotic region $\partial H$ and $\partial \theta$
cannot be extended to conserved currents but their leftmoving
linear combination $\partial \phi$ is exactly conserved.  The
current which asymptotically looks like
$(j^\theta (z), \t j^\theta (\zb)) = {i R \over 2}(\partial \theta, \bar \partial \t \theta)  $ is conserved in the full
theory and the corresponding charge $P^\theta$ is the momentum around the
cigar in units of the inverse radius.  Since it is not purely leftmoving, the winding number is not conserved.

We conclude that the exact theory on the cigar has three conserved currents. $\partial \phi$ is leftmoving, $\bar \partial \t \phi$ is
rightmoving and $j^\theta$ has both a leftmoving and a
rightmoving component.  In the asymptotic region where the theory
looks like a linear dilaton theory one can describe the operators
in terms of $H$, $\t H$, $\theta$ and $\t \theta$.  But in
the full theory it is better to use $\phi \approx H+Q\theta$ and
$\t \phi \approx \t H+Q \t \theta $. 

When $\IR^{d-1,1}$ is tensored with the cigar, the currents $T_d,G^{\pm}_d,J_d$ of the $d$ free bosons and fermions should be added to this CFT. All the conserved chiral currents now have additional pieces from the free CFT. The momentum around the cigar as defined above is still a symmetry, and there are also $d$ conserved momenta in the flat directions. We can add the superconformal ghosts to this theory to form a consistent string background. The central charge of this theory, $\hat c=2(1+Q^2)+d$ should be set to $10$ which gives $Q=\sqrt{\half (8-d)}$.

To construct theories with spacetime supersymmetry, we need to introduce supercharge operators. As seen above, the $U(1)$ $R$ current of the worldsheet algebra in these theories involves the compact boson in addition to the worldsheet fermions. This means that the standard R-NS construction of the theory yields a spectrum which does not have a good spacetime interpretation as particles propagating in the $d+2$ dimensional curved spacetime; the spin of a particle and its momentum around the cigar are not independent. Rather, the theories have a natural holographic interpretation as a non-gravitational theory living in $d$ dimensions. As we shall see, this feature is intimately related to the structure of supersymmetry in such curved spacetimes. 

To emphasize the above point, we shall first construct the bosonic Type $0$ theories, and then generate the supersymmetric Type II theories as a $\IZ_2$ orbifold of the former. The bosonic Type $0$ theories have no surprises and we shall quickly review the standard construction. We shall then see that the chiral GSO projection to get to the Type II theories has a subtlety involved, and the spectrum looks slightly unusual from the cigar perspective. We shall then present the classification of the particles of the Type II theories  as off shell operators in the $d$ dimensional holographic theory.

\subsec{Type 0 theories on the cigar $\times \IR^d$}
We first construct the chiral (left or right moving) part of the vertex operators by demanding that they are physical operators in the theory. For example, in the pure cigar ($d=0$) case, the the form of the vertex operators in the asymptotic region where all the worldsheet fields are free is (ignoring discrete states)
 \eqn\htherh{\eqalign{
 &\CO_{nkp}^{NS}=e^{-\varphi+in H + ik\theta + (-1+p)\rho} \cr
 &\CO_{nkp}^R=e^{-\half\varphi+ i(n+\half) H + ik\theta +
 (-1+p)\rho}}}
Here $n\in Z$ and $p\ge 0$, and $\varphi$ is the bosonized superconformal ghost \fms . In the higher dimensional theories, the operators are functions of the free fields as well. Physical operators also obey the condition of BRST invariance. To form the closed string theories, we put together the left and right movers with the following conditions: 
\item{1.} Modular invariance demands a diagonal GSO projection, i.e. the same boundary conditions for the left and rightmoving fermions, together with
\eqn\typeOtwod{\eqalign{
  0B: & \;(-)^{j_L} = (-)^{j_R}; \cr
  0A: & \; (-)^{j_L} = (-)^{j_R} \; {\rm in}\; NS, \cr
      & \; (-)^{j_L} = (-)^{j_R+1} \; {\rm in}\; R. \cr
}}
where $j_{L,R}$ are the leftmoving and rightmoving fermion number currents. 
\item{2.} The parity even combination of the lowest NS-NS winding operators $\CO^{NS}_{n=0,k=\pm\half}, \tilde \CO^{NS}_{n=0,k=\mp \half}$ is the interaction term in the sine-Liouville theory written in the dual variables. The duality between the two theories shows that the spectrum must have the above two NS vertex operators. Imposing locality of the rest of the spectrum relative to these vertex operators, demanding that the Liouville momentum of $\rho$ and $\tilde \rho$ must be the same, and the level matching condition determines the spectrum. 

At this point, we should make a remark concerning the nature of these operators. In theories which asymptote to linear dilaton backgrounds, there is no state-operator correspondence. Non-normalizable modes correspond to local operators in the theory, and the normalizable modes are the states, or vacuum deformations in spacetime \seiberg . For the non-critical superstring theories with $d \le 4$, the sine-Liouville interaction is a non-normalizable local operator and can be put in the action. The cigar metric is normalizable for all the theories. For $d > 4$, the sine-Liouville operator is normalizable as well. We shall not grapple with this issue in the following, and shall study all the theories, including $d=6$. 

The Type $0$ theories as defined above for all $d$ have a straightforward interpretation in terms of the cigar geometry. The spectrum can be classified as a set of particles with increasing masses. Asymptotically, all the propagating modes are determined by a free field propagating on the geometry, so that the particles have all integer momenta and winding (which is not a good quantum number) allowed by the equation of motion. 

The lowest lying modes are a tachyonic scalar, and a graviton multiplet with $d^2$ degrees of freedom in the NS-NS sector, and a set of massless R-R fields appropriate to the particular dimension\foot{For $d=0$, there are no transverse oscillators and no graviton, there are only a few field theoretic states in the spectrum.}. The winding modes in the R-R sector have a possible interpretation as Wilson lines of R-R potentials around the tube. This interpretation is only valid asymptotically, where the field strength vanishes. The type $0$ spectra are presented in Appendix A with an example.

We can study the high energy behaviour of these theories and get an estimate for the asymptotic density of states by a saddle point approximation of the partition sum. The result for the mass density of states as a function of the spacetime dimension $d$ is (Appendix B):
\eqn\massdos{
 \rho(m) \sim m^{-(d+2)}\exp({m\over m_0}), \qquad m_0 = (\pi\sqrt{d\alpha'})
}
As noted in \kutsei , \kutasov , this is unlike compactification to $d$ dimensions wherein the string at high energies does see all the ten dimensions. These string theories are in this sense, truly $d$ dimensional.

\subsec{The chiral GSO projection and Type II theories}

The symmetries of the 0A and 0B theories are the momentum around the cigar, and naively two (vector and axial) $U(1)$ R symmetries on the worldsheet. Only the first one under which all the R-R fields pick up a negative sign and is the one used in the type $0$ projection, is a true symmetry of the theory. This is clear from the sine-Liouville interaction $\CL^{SL}_{int} = \psi \t \psi \, e^{-{1 \over Q} (\rho +\t \rho + i (\theta - \t \theta ))} + c.c$ where $\psi = \psi_\rho + i \psi_\theta$ is the superpartner of $\rho + i \theta$ and $\t \psi$ is its rightmoving counterpart. The chiral rotation of the fermions is not a symmetry, the rotation of the left and right moving fermions in opposite directions is. It is also clear that there {\it is} a conserved chiral $U(1)$ current which rotates the left moving fermion by an angle $\alpha$ and simultaneously translates the left moving boson $\theta$ by $Q\alpha$.

From the cigar point of view, the non-conservation of the chiral rotation  can be understood as due to the anomaly at one-loop in the $U(1)$ current $j_L$ which rotates only the left moving fermions caused by the curvature of the cigar:
\eqn\anomaly{
 \p_\alpha j_L^\alpha = R (\epsilon^{\alpha\beta}\,\p_\alpha\rho\,\p_\beta\theta) ,
}
where $R = -2 D^a D_a \Phi = {-Q^2 \over 2\cosh^2{Q\rho\over 2}}$ is the Ricci curvature of the cigar. 

Due to the special form of the curvature in two dimensions, we can define a new current which {\it is} conserved. Changing to complex coordinates on the worldsheet, this current is the sum of the chiral rotation and another piece proportional to the left moving momentum:
\eqn\conscurr{
 \pb j_G := \pb (j_L + Q (\tanh{Q\rho\over 2}) \p \theta) = 0
}
which reduces to the $U(1)$ $R$ current of the  $\CN=2$ SCFT \ntwowss\ in the asymptotic region. 

We conclude that to perform a chiral $\IZ_2$ projection to get the type II theories, we {\it must} use the $\IZ_2$ symmetry generated by the conserved current above, which acts in the asymptotic region as 
\eqn\defG{
 G = (-)^{j_L+Q k_L}.
}

\medskip

This GSO projection is implemented by introducing the target space supercharge in the twisted sector, as in \kutsei. For example, in the pure cigar case, we demand that the OPE of the $(1,0)$ operator
 \eqn\defso{S=e^{-{\varphi \over 2} + i{\phi\over 2}
}}
with the physical operators is local. When $\IR^{d}$ is tensored to this background, $S=e^{-{\varphi \over
2} +i{\phi\over 2}}$ and $\bar S=e^{-{\varphi \over 2} -
i{\phi\over 2}}$ are each multiplied by spin fields which are
spinors of $Spin(d)$ (for $d/2$ even these are conjugate spinors
and for $d/2$ odd they are the same spinor). There is also a similar condition on the rightmoving side depending on whether the theory is IIA or IIB.

The algebra of the supercharges can be deduced by examining the currents in the asymptotic region. For $d/2$ odd or even, one has respectively 
\eqn\susyalg{
\{\CS_\alpha,\bar \CS_\beta\}= 2\gamma^\mu_{\alpha\beta}P_\mu, \qquad {\rm or} \qquad 
 \{\CS_\alpha,\bar \CS_{\dot \beta}\}= 2\gamma^\mu_{\alpha\dot\beta} P_\mu.
}
Note, in particular that it does not contain translation in $\theta$. In fact, the symmetry generator $P^\theta$ corresponding to the translation around the cigar is an $R$ symmetry.
\eqn\pssbaralg{
  [P^\theta, \CS_\alpha]=\half \CS_\alpha, \qquad [P^\theta, {\bar \CS}_{\dot \alpha}]=-\half {\bar \CS}_{\dot \alpha}. 
}

This means that the fields in a given supersymmetry multiplet do not all have the same value of momentum. Thus, the allowed momenta around the cigar of a particle is correlated with its behaviour under Lorentz rotations in spacetime. For instance, in the zero winding sector, the graviton has only even momenta, and the tachyon\foot{The tachyonic zero mode is projected out and the field is no longer tachyonic. The type II theories are stable.} has only odd momenta. The fermionic states which arise in the twisted sector have half integer momenta. This means that they are antiperiodic around the tube. This is the expected behaviour for spinors which are single valued on the cigar (the spin structure which can be extended). We find also a discrete winding symmetry which is not natural from the cigar perspective. This discrete symmetry has its most natural interpretation from the Calabi-Yau or NS5-brane description of the theories, as described later in section 4. The type II spectrum is presented in Appendix A with an example.

In the Green-Schwarz formulation of the superstring, the states of the theory are in a representation of the zero modes of the spinors which are the superpartners of the bosonic fields. The quantization is treated in Appendix C. Here we point out two features of these theories. The first one is how the above non-smooth spacetime structure manifests itself in the Green-Schwarz formalism. In this description, the conserved current which corresponds to momentum around the cigar is really a combination of the naive momentum and a piece which acts on the worldsheet fermions. The spin of a particle and the momentum around the cigar are thus not independent of each other. 

The second issue regards the comparision to the 10-dimensional Green-Schwarz string. In that case, in the light-cone gauge, the symmetry group $SO(8)$ has the property of triality relating the chiral spinor and vector representations which is crucial in showing that the spectrum is the same as that obtained by the R-NS formalism. In our theories in lower dimensions, there is no triality, but correspondingly, the NS-NS ground state is not a graviton built out of left and right moving vectors, rather it is the scalar (tachyon) field with one unit of momentum. There is no need for triality in these theories. 

\subsec{Holographic interpretation}

The bosonic spectrum of the supersymmetric theories cannot be organized as multiplets of the $d+2$ dimensional bosonic Poincare group, but the full spectrum {\it can} be organized into multiplets of spacetime supersymmetry. This algebra \susyalg\ is effectively $d$ dimensional and the modes of the $d+2$ dimensional fields are naturally classified as $d$ dimensional off-shell currents (operators). This is consistent with the holographic interpretation of such theories \dilholo . We shall illustrate this using the $d=4$ example.

The spacetime supersymmetry algebra of the theory is a four dimensional $\CN=2$ algebra with a $U(1)$ R charge which we identified as the momentum around the cigar  $R=2 P^\theta$. The six dimensional fields arrange themselves into multiplets of this supersymmetry. The Kaluza-Klein modes with a certain value of momentum around the cigar will be on-shell particles in five dimensions, and so they will fall into the current representations of the four dimensional algebra [{\it e.g.} Table 1,2 below]. These off shell four dimensional currents are conserved because of the gauge invariance in five dimensions. 

\medskip
\centerline{Table 1: Tachyon multiplet}
\medskip
\centerline{\vbox{\offinterlineskip
\hrule
\halign{\vrule # & \strut\ \hfil #\ \hfil & \vrule # & \ $#$ \ & \vrule # & \ $#$ \ & \vrule # & \ $#$ \ & \vrule # & \ # \ & \vrule #  \cr
height3pt&\omit&&&&&&&&&\cr 
& 6d rep:&&(A_a)_{\rm RR}&&(\psi_1,\psi_2),({\bar \psi_1},{\bar \psi_2}) && (T,T^*),(T_w,T^*_w) && Tachyon &\cr
& 4d rep:&&(V_\mu, D) && (\psi_1,\psi_2),({\bar \psi_1},{\bar \psi_2}) && (\phi_1,\phi_1^*),(\phi_2,\phi_2^*)&&$\CN=2$ vector current& \cr
& $n = \half R$ : &&0 && (\half,\half),(-\half,-\half) &&  (1,-1,),(0,0)&&& \cr
height3pt&\omit &&&&&&&&&\cr }
\hrule
}}
\bigskip

\centerline{Table 2: Graviton multiplet}
\medskip
\centerline{\vbox{\offinterlineskip
\hrule
\halign{\vrule # & \strut\ \hfil #\ \hfil & \vrule # & \ $#$ \ & \vrule # & \ $#$ \ & \vrule # & \ $#$ \ & \vrule # & \ # \ & \vrule #  \cr
height3pt&\omit &&&&&&&&&\cr 
& 6d rep:&&(G,B,\phi)_{\rm NSNS} &&(\chi,\psi_\mu),({\bar \chi},{\bar \psi_\mu}) &&(F_{\pm\mu},F^{(\pm)}_{\mu\nu})_{\rm RR}  && Graviton& \cr
& 4d rep:&&(T_{\mu\nu},B_{\mu\nu},A_\mu,B_\mu) &&(\chi,\psi_\mu),({\bar \chi},{\bar \psi_\mu})&&(\p_\mu D \pm k_\nu^2 V_\mu), (V^{(\pm)}_{\mu\nu}) && $\CN=2$ supercurrent&\cr
&$n= \half R$: &&0 && (\half,\half),(-\half,-\half)&& (1,-1),(0,0) &&& \cr
height3pt&\omit &&&&&&&&&\cr }
\hrule
}}
\medskip

\leftskip=35pt
\rightskip=35pt
\baselineskip=12pt

\ndt {\it Supersymmetry structure of the $d=4$ spectrum:} These are the first two Kaluza-Klein modes of the six dimensional fields in the type IIA theory. In six dimensions, these are the tachyon multiplet and the graviton multiplet. They are classified by their properties under the 6d Poincare algebra and the 4d SuperPoincare algebra. The momentum around the cigar is proportional to the U(1) R charge. The spinors are two component spinors in ${\bf 2}$ and ${\bf \bar 2}$ of $SO(4)$.

\leftskip=0pt
\rightskip=0pt
\baselineskip=18pt

\bigskip

The details of the various type II theories differ slightly from each other. Some of the higher dimensional theories have been studied in \refs{\bilal , \eguchi , \mizo }. These authors however, were interested in constructing modular invariant partition functions in particular cases, and the spacetime picture is not dwelt upon. In Appendix B, we present the modular invariant partition functions for the perturbative string spectrum for the various values of $d$. These are constructed by the standard technique \polchinski\ of counting the physical momentum and winding forced by the above GSO projection in the light cone gauge. In the next section, we shall visit the various theories and highlight the interesting features that each of them have.

\newsec{Special features of the various theories}

\subsec{$d=0$}
In the theory of the pure cigar, the spectrum looks smooth in that it can be interpreted as a set of particles propagating on the cigar\foot{This is loose usage of language - in two dimensions, there is no notion of a particle.}. There are no transverse oscillators, and hence no infinite tower of states that we find in higher dimensional theories. We can explicitly write down the all the vertex operators easily in this case. In the IIB theory, we have (all operators with $p=|k|$):
\eqn\iibspectrum{\eqalign{
 &e^{-\varphi-\tilde\varphi+i(N+\half)(\theta-\tilde \theta)+(-1 +
 |N+\half|)(\rho+\tilde \rho)} \qquad\qquad N=0,\pm 1,\pm 2,...\cr
 &e^{-\varphi-\half\tilde\varphi -\half i \tilde H
 -i(N+\half)(\theta+\tilde \theta)+(-\half +N)(\rho+\tilde \rho)}
 \qquad\qquad N=0,1,2...\cr
 &e^{-\half\varphi-\tilde\varphi -\half i H
 -i(N+\half)(\theta+\tilde \theta)+(-\half +N)(\rho+\tilde \rho)}
 \qquad\qquad N=0,1,2...\cr
 &e^{-\half\varphi-\half\tilde\varphi +\half i(H+\tilde H)
 +i N(\theta+\tilde \theta)+(-1 +N)(\rho+\tilde \rho)}
 \qquad\qquad N=0,1,2...\cr}}
In the NS-NS sector we have only winding modes of the scalar. In the NS-R and the R-NS sector, we find spacetime fermions.  As can be seen from the $H$ and $\tilde H$ behavior of the vertex operators they both have spin $-\half$. As mentioned earlier, the half integer momentum is the expected behavior for spacetime spinors which are single valued on the cigar. This agrees with the interpretation of a black hole at finite temperature. From the R-R sector we find a
spacetime boson which is periodic around the tube.  We interpret this boson as the R-R scalar of the IIB theory.  Note that the two fermions have only negative $k$ and the R-R scalar has only positive $k$.

\ndt In the IIA theory we find
 \eqn\iiaspectrum{\eqalign{
 &e^{-\varphi-\tilde\varphi+i(N+\half)(\theta-\tilde \theta)+(-1 +
 |N+\half|)(\rho+\tilde \rho)} \qquad\qquad N=0,\pm 1,\pm 2,...\cr
 &e^{-\varphi-\half\tilde\varphi +\half i \tilde H
 +i(N+\half)(\theta+\tilde \theta)+(-\half +N)(\rho+\tilde \rho)}
 \qquad\qquad N=0,1,2...\cr
 &e^{-\half\varphi-\tilde\varphi -\half i H
 -i(N+\half)(\theta+\tilde \theta)+(-\half +N)(\rho+\tilde \rho)}
 \qquad\qquad N=0,1,2...\cr
 &e^{-\half\varphi-\half\tilde\varphi +\half i(-H+\tilde H)
 -i (N+\half)(\theta-\tilde \theta)+(-\half +N)(\rho+\tilde \rho)}
 \qquad\qquad N=0,1,2...\cr
 &e^{-\half\varphi-\half\tilde\varphi +\half i(H-\tilde H)
 +i N(\theta-\tilde \theta)+(-1 +N)(\rho+\tilde \rho)}
 \qquad\qquad N=0,1,2...\cr}}
We again find only winding modes of the scalar in the NS-NS sector. The NS-R
and R-NS sectors lead to antiperiodic spacetime fermions. 
Their spins are $+\half$ and $-\half$, which is consistent with the 
spacetime parity of the IIA theory. The R-R sector leads to winding modes which have an asymptotic interpretation as Wilson lines of the R-R one form around the tube.

One aspect worth mentioning about this theory is that the physical spectrum is extremely constrained due to the lack of transverse oscillators. There are no gravitons, and the supercharges are not part of the spectrum. In fact, to obtain a modular invariant partition function[Appendix B], we cannot impose the Dirac equation, and we need to demand it as a further condition to read off the physical spectrum. Even when we do this, the bosons and fermions are not paired and so the partition function does not vanish.

\subsec{$d=2$}

In the theories with $d=2,6$, there is a further subtlety in the chiral projection which arises because of the way the conserved fermion number current $j_G=j_L+Q k_L$ acts in these theories. The symmetry $G=e^{\pi i j_G}$ has a $\IZ_4$ (and not $\IZ_2$) action on the fields of the type 0 theories. The symmetry $G_1=(-)^{2j_G}$ is a  $\IZ_2$ subgroup of the $\IZ_4$ symmetry in the type 0 theories. Orbifolding by $G_1$ gives another bosonic theory which we call the type $0'$ theories. These theories has a $\IZ_4/ \IZ_2=\IZ_2$ symmetry $G_2$ by which they can be further orbifolded to get the supersymmetric type II theory. 

In the $d=2$ dimensional case, $Q=\rtt$ which gives $R={2\over \rtt}$. The $\IZ_2$ symmetry in the type 0 theories is $G_1=(-)^{2j_L+2\rtt k_L}$. The NS-NS operators have charge $(-)^{2\rtt k_L}$ and the R-R operators have charge $(-)^{2\rtt k_L+1} $ under this symmetry (This is because there are only two spin fields in four dimensions). The construction of the type $0'$ theories is implemented by demanding the presence of the $R\pm R\mp$ operators in the 0'B and $R\pm R\pm$ operators in the 0'A with winding $w=\pm\half$. The chiral $\IZ_2$ symmetry of these theories is $(-)^{j_L+\rtt k_L}$, orbifolding by which we get the type II theories. This is implemented as usual by demanding the supercharges to be present in the spectrum. The supercharges in this theory have $(n,w)=\pm(\half,\pm{3\over 4})$.

\subsec{$d=4$}

We have been presenting examples from the $d=4$ theory above to illustrate the various general features, and shall only make one remark here. The asymptotic radius of the cigar is the self-dual radius. Asymptotically, the current $\p \theta$ is a physical current, but the currents $e^{\pm i \irt2 \theta}$ are not. The $SU(2)$ symmetry of the free boson is therefore not present in this model. 

\subsec{$d=6$ or $k=2$ NS5-branes}
\ndt This example is particularly rich because of the explicit geometric description in terms of NS5-branes referred to in the introduction. We shall focus on three main points:
\item{1.} The equivalence between the cigar and sine-Liouville theories. We shall see that the two interactions are related by a symmetry.
\item{2.} The description of the moduli space of the theory, and 
\item{3.} A discussion of the discrete $\IZ_2$ symmetries present in the various 5-brane and cigar theories. 

\ndt A quick review of the picture we shall use is the following: The conformal field theory in the asymptotic region is that of two parallel NS5-branes. This can be explicitly seen by fermionizing the angular coordinate $\theta$ on the tube to two free fermions $e^{\pm i\theta} \equiv \irt2 (\psi_1 \pm i\psi_2)$, which along with the fermion $\psi_\theta \equiv \psi_3$ generates a left moving $SU(2)_2$.\foot{In this case, $Q=1 \Rightarrow R=2$ which is the free fermion radius.} The theory has an $SU(2)_L \times SU(2)_R = SO(4)$ symmetry generated on the worldsheet by the rotation of the three leftmoving and three rightmoving fermions. The worldsheet interaction which resolves the singularity at the origin breaks the symmetry down to a global diagonal $SO(3)$ corresponding to the rotation of the three directions transverse to the two separated 5-branes. 

The relationship between the cigar and sine-Liouville theories is a
strong-weak coupling duality on the worldsheet. For large $k$, the
cigar has small curvature and the classical description in terms of the metric is a good approximation to the full theory. The sine-Liouville term decays much faster than the cigar metric asymptotically, and the sine-Liouville lagrangian is strongly coupled. For small $k$ on the other hand, the sine-Liouville term asymptotically dominates over the cigar metric which has large curvature, and it is the sine-Liouville which is a better description in terms of a weakly coupled worldsheet theory. For general $k$, the effective lagrangian in the full quantum theory has both these terms, and the dominance of one of these terms over the other is governed by the value of $k$. This has been confirmed by explicit calculations of scattering amplitudes [{\it e.g.} \kutgivads\ and refs. therein] where one sees poles corresponding to both the terms which can be used to compute an explicit relation between the two couplings. 

In our theory with $k=2$, both the terms decay at the same rate. As
we shall see, they are related by a rotation in the $SO(3)$ symmetry group mentioned above. This gives an example where the kinematic structure determines explicitly that both the terms are present in the lagrangian and also determines the relation between the strength of the two couplings. 

The three $\CN=2$ invariant currents $\psi_i e^{-\rho}$ (in the $-1$ picture) are in a triplet under $SU(2)_L$. The meaning of these currents is better understood when expressed in the $0$ picture in the variables of the cigar: $(\psi_\rho \mp i\psi_\theta)e^{-\rho \pm i\theta}, (\psi_\rho \psi_\theta  + \p \theta)e^{-\rho}$ - the first two terms are nothing but the sine-Liouville interaction, and the third term is the first order correction from the cylinder towards the cigar metric. We can express these currents in a manifestly $SU(2)$ covariant manner as the fermion bilinears $A_i= (\psi_\rho\psi_i - \half \epsilon_{ijk}\psi_j\psi_k)e^{-\rho}$. There are also the corresponding rightmovers. Noting the Clebsch-Gordon coefficients relating the $\bf 4$ of the $SO(4)$ and the $\bf 3$'s  of the left and the right $SU(2)$ (the 'tHooft symbols),
\eqn\cgcoeff{
\eta^i_{\mu\nu}, {\t \eta}^i_{\mu\nu} = \delta^0_{[ \mu} \delta^i_{\nu ]} \pm \half \epsilon^i_{jk} \delta^j_{\mu} \delta^k_{\nu},
}
\ndt we can write down the proposed interaction for the theory of the non-coincident parallel NS5-branes. The matrix $\CL_{\mu\nu} = \eta^{i\,\sigma}_{\mu} {\t \eta}^j_{\sigma\nu} A_i \t A_j$ is in the symmetric traceless ($\bf 9$) of the $SO(4)$, and the interaction 
\eqn\covint{
S_{int} = \int d^2 z \;X^\mu X^\nu \CL_{\mu\nu}
}
corresponds to separating the branes in the direction $X^\mu$ with
the center of mass at the origin. A choice of $X^\mu$ breaks the $SO(4)$ symmetry to an $SO(3)$ rotation symmetry. Changing the position of the branes by a $SO(4)$ transformation leads to a different $SO(3)$ being preserved. This is reflected in the lagrangian by simultaneously conjugating the matrix $\CL_{\mu\nu}$ by the same transformation. In keeping with the conventions used in the earlier sections, we shall relabel the four transverse directions as $\mu=6,..9$.

Since the two 5-branes are identical, the moduli space of the theory is $\IR^4/\IZ_2$, parameterized above by $X^\mu$. In the $SL_2(\IR)/U(1)$ description of the theory, there are only four independent $(1,1)$ physical states. Two of the three chiral operators $A_i$ (and correspondingly the rightmovers) are related by the spectral flow operation by one unit in the $SL_2/U(1)$ theory. This means that only two of them are independent states. The wavefunction of a state is in general a linear combination of the wavefunctions of the states reached by spectral flow from a given state. The coefficients of these wavefunctions depends on the given point in moduli space, because this determines the boundary conditions at the tip of the cigar \refs{\maldoog ,\juan }. The above four states are in the representation ${\bf 1} + {\bf 3}$ of the $SO(3)$, corresponding to the radial motion of the branes and the motion on the three-sphere of given radius. 

In the subspace that preserves the momentum symmetry of the cigar and the parity of the angular coordinate, there is only one exactly marginal deformation. This was interpreted in \horikap\ as the metric deformation. As mentioned above, this interpretation is only correct for large $k$ where it dominates - as we see here, the spectral flow relation means that the deformation is a combination of operators as given by \covint .

In the lagrangian \covint\ proposed above, the fact that there are
only four independent operators is seen as a condition of second order conformal invariance of the theory. In the linear dilaton background, all the nine operators are conformal. In the background given by the interaction corresponding to say $X^6 \ne 0, X^{7,8,9}=0$, only the operators $\CL_{6\mu}$ remain conformal.  To determine exactly marginal deformations, we add an arbitrary combination of the nine operators $C_{\mu\nu} \CL^{\mu\nu}$ to the free action, and find that the theory remains conformal iff $C_{\mu\nu}=x_\mu x_\nu$ [Appendix D] which is exactly the four dimensional moduli space written above. 

To summarize, there is no pure sine-Liouville or pure cigar theory, only particular linear combinations given by \covint\ are consistent with the $SO(4)$ being broken to an $SO(3)$ rotation symmetry. The theory has a four dimensional moduli space $\IR^4/\IZ_2$. 

We now make a few remarks:
\item{1.} The discrete winding symmetry in the theory is identified in the NS5-brane picture as the rotation by $\pi$ in a plane containing the two branes, exchanging the two. This $\IZ_2$ symmetry completes the $SO(3)$ into an $O(3)$. 
\item{2.} The separation of the branes means that the origin is not singular, and this gives a unique spin structure in the transverse space. This forces the fermions to pick up a phase of $\pm i$ under the action of the exchange symmetry generator, precisely as seen in the spectrum. The fermions also transform in the $SU(2)$ cover of the global $SO(3)$.
\item{3.}  We can make precise the relation of the cigar and 5-brane theories defined by the various GSO projections.  The type $0'$ and II theories on the cigar correspond to the type $0$ and II theories on the two 5-branes. The $\IZ_2$ orbifold by the exchange symmetry mentioned above takes us from the type $0'$ to the type $0$ theory on the cigar. The smooth (but tachyonic) type 0 theory on the cigar is identified with the theory of one 5-brane on $ \IR^4/\IZ_2$ away from the origin. This theory does not have the $SO(3)$ symmetry. 
\item{4.} To restate a point, the orbifold which gives the 5-brane theories from the Type 0 theory on the cigar does not have a geometric interpretation like changing the radius. The type $0A(B)$ and Type IIA(B) theory on the 5-branes can be obtained by quotienting by a chiral $\IZ_2$ either the type $0A(B)$ theory with radius $R=2$, or the type $0B(A)$ with radius $R=1$ (which is produced from the former $0B(A)$ by a geometric quotient).

\subsec{A note on the $\CN=(4,4)$ algebra in the $d=6$ theory.}
This small subsection is slightly outside the main flow of the paper. The main points in this subsection are the existence of an extended superconformal algebra in the $d=6$ theory, and a comment on the related $D1/D5$ system.

Consider first the theory of $k=2$ coincident NS5-branes. In a physical gauge where the lightcone directions along the NS5-branes are fixed, the degrees of freedom on the worldsheet are four free bosons (parallel to the branes) + four free fermions which have an $\CN=(4,4)$ superconformal structure with $c=6$, and a linear dilaton ($\rho$) + four free fermions ($\psi_\mu= (\psi_\rho,\psi_i)$). There is also an $\CN=(4,4)$ superconformal algebra involving the latter fields \chs\ of central charge $c=6$, which extends our algebra \ntwowss . There are four fermionic currents and three R currents forming an $SU(2)_1$. The currents of the algebra are: 
\eqn\extscalg{\eqalign{
 &T = -\half (\partial \rho)^2 - \half \psi_\mu\partial \psi_\mu 
 - \half \partial^2\rho \cr
 &G_\mu = {1 \over \rt2} \psi_\mu \partial \rho + {1\over 6 \rt2} \epsilon_{\mu\nu\sigma\tau} \psi^\nu \psi^\sigma \psi^\tau + {1 \over \rt2} \p \psi_\mu  \cr
 &J_i = \half (\psi_\rho \psi_i +\half \epsilon_{ijk} \psi_j \psi_k) = \half \eta^{\mu\nu}_i\psi_\mu \psi_\nu.
}}

As shown above, when the branes are noncoincident, the moduli are parameterized by the operators $\CL_{\mu\nu}$ constructed using the chiral operators $A_i = (\psi_\rho \psi_i  - \half \epsilon_{ijk} \psi_j \psi_k)e^{-\rho} \equiv {\bar J_i} e^{-\rho}$ and their right moving counterparts. These operators commute with the above algebra and the $\CN=(4,4)$ superconformal symmetry is preserved at all points in the moduli space. This is consistent with the fact that there are $16$ supercharges in spacetime even when the two branes are non-coincident (but parallel). The $SO(4)=SU(2) \times SU(2)$ of the $J_i, \t J_i$ is identified with the rotation of the four directions parallel to the brane. 

A fact worth pointing out is that for a system of $k$ NS5-branes, in addition to the above algebra \extscalg\ with $c=6$, there is a second $\CN=4$ algebra with $c=6(k-1)$. The $R$ currents of this algebra contains $\bar J_i$ and another piece from the bosonic $SU(2)_{k-2}$, and the slope of the dilaton is different. Both these algebras have the same algebraic structure (called the `small' $\CN=4$ algebra), and they are both subalgebras of the `large' $\CN=4$ algebra. This `large' algebra contains both the sets of $SU(2)$ $R$ currents and its generators do not contain the improvement terms from the linear dilaton \refs{\chs,\seiwitt}. 

As explained in \seiwitt , these two conformal field theories are spacetime descriptions of the short and long string excitations respectively in the background of $k$ NS5-branes and many fundamental strings.\foot{In the gauge we have chosen, at distances where we are in the near horizon limit of the 5-branes but not near horizon of the strings, the theory of short strings in this background is the same as the one we have been studying with 5-branes alone \kutgivpelc .} Equivalently, they describe the Coulomb and Higgs branches respectively of the gauge theory on a D1-brane in the presence of D1-branes and D5-branes in the IIB theory. In general, the slope of the dilaton in the Higgs branch differs in sign and magnitude from the Coulomb branch, and the symmetries of the two branches are different. 

For the case $k=2$, there is a symmetry mapping one theory to the other. In this special case, both the chiral subalgebras have $c=6$, and the string couplings are inverses of each other. There are two $SO(4)=SU(2)_L \times SU(2)_R$ - one corresponding to the rotation of the four directions parallel to the 5-brane and the other to the rotation of the four transverse directions. In the Coulomb branch, the parallel $SO(4)$ generated by the $J_i, \t J_i$ is preserved and the transverse $SO(4)$ is broken to a global diagonal $SU(2)$ by turning on the moduli parameterized by $\bar J^i, \bar  {\t J^i}$ corresponding to the separation of the branes. In the Higgs branch, the roles of $J^i$'s and $\bar J^i$'s are reversed (in this system, the moduli are the self dual NSNS B field and a linear combination of the RR zero form and four form). Given the identification of the long tube of the two branches as in \seiwitt , one can summarize the above by saying that there is a symmetry relating the short and long string systems given by: $(\rho,\psi_\rho) \rightarrow - (\rho,\psi_\rho)$. Starting from the weak coupling end of either theory, turning on the moduli to a non-zero value caps off the infinite tube to a semi-infinite cigar.

\newsec{The global symmetries of the various theories}

\ndt The global symmetries of the various theories are the following:

\centerline{Table 3: {\it Global Symmetry structure}}
\medskip
\centerline{\vbox{\offinterlineskip
\hrule
\halign{\vrule # & \strut\ \hfil #\ \hfil & \vrule # & \ \hfil $#$ \hfil \ & \vrule # & \ \hfil $#$ \hfil \ & \vrule # & \ \hfil $#$ \hfil \ & \vrule # & \ \hfil $#$ \hfil \ & \vrule #\cr
height3pt&\omit&&&&&&&&&\cr 
&{\bf Theory}&&  d=6 && d=4 && d=2 && d=0  & \cr
height3pt&\omit&&&&&&&&&\cr 
\noalign{\hrule}
height3pt&\omit&&&&&&&&&\cr 
& {\bf Supersymmetry} && \CN=(2,0)\; $or$\; \CN = (1,1)  && \CN = 2 && \CN = (4,0) \; $or$ \; \CN = (2,2)  && \CN=2  &\cr
height3pt&\omit&&&&&&&&&\cr 
\noalign{\hrule}
height3pt&\omit&&&&&&&&&\cr 
& {\bf Bosonic Symmetry} && SO(5,1) \times O(3)_R && SO(3,1) \times U(1)_R && SO(1,1) \times (U(1) \times \IZ_2)_R && U(1)_R  &\cr
height3pt&\omit&&&&&&&&&\cr}
\hrule
}}

\medskip
\leftskip=35pt
\rightskip=35pt
\baselineskip=12pt

\ndt Only the Lorentz part of the full Poincare group in the flat directions is written above. The supersymmetry algebra could be chiral or non-chiral in $d=6,2$ depending on if the theory is IIA or IIB. The fermions in the theories with $d=6,4,2$ transform in the $Spin(d-1,1)$ in the flat directions. They also transform in the spin cover of the $R$ symmetry group: In $d=6$, they are charged under $SU(2)$ of rotation, they have half integer charge under the $U(1)$'s  in the $d=4,2,0$ theories, and they pick up a phase of $\pm i$ under the $\IZ_2$ generator in the $d=6,2$ theories.  

\leftskip=0pt
\rightskip=0pt
\baselineskip=18pt

\bigskip

From the CFT description of the non-crtitical string theories, we saw in section 2 the appearance of all these symmetries. From the spacetime point of view, the super-Poincare group in the flat directions is natural. As we saw earlier, the $U(1)$ part of the $R$ symmetry group is interpreted as the conserved momentum around the cigar. In the previous section, we saw using the NS5-brane picture, exactly how all the symmetries of the $d=6$ theory arise. The full $R$ symmetry groups for all the theories including the discrete $\IZ_2$ symmetries are naturally seen in the singular Calabi-Yau or NS5-brane description, which is what the rest of this section is devoted to. 

As mentioned in the introduction, all the non-critical theories above are conjectured to arise as near horizon geometries of wrapped NS5-branes, which are dual to string theory near the singularity of certain singular Calabi-Yau manifolds tensored with flat space in a double scaling limit \refs{\kutgiv,\kutgivpelc,\oogvafa }. The global symmetries of our theories are nothing but the symmetries of these brane configurations in the near-horizon limit, or equivalently, those of the dual geometry whose action remains non-trivial in the double scaling limit. We will now describe the T-duality between the singular spaces and wrapped NS5-branes [\mukhi , and refs. therein] and their respective deformations, and track which symmetries survive in the scaling limit. A linear sigma model analysis of this T-duality has been done in \tong .

The non-critical theory in $d$ dimensions is conjectured to be equivalent to string theory on $\IR^{d-1,1} \times M^{10-d}$ where $M^{10-d}$ is defined as the hypersurface $\sum_{i=1}^n z_i^2 = \mu$ in $\IC^{n}$ where $n=(12-d)/2$.\foot{For $d=0$, we should consider the Euclidean theory, and in terms of the brane picture, it is the theory of an Euclidean NS5-brane completely wrapped on a non-compact Calabi-Yau 3-fold embedded in $\IR^8$.} By considering as above non-compact geometries, we have already taken the limit in which we zoom in on the singularity of the Calabi-Yau. In terms of the dual 5-brane description which we present below, the branes are stretched out to infinity even in the non-flat directions, describing deformed intersecting surfaces in these directions for $d \le 4$. The only global symmetries left are the ones involving the directions fully transverse to the brane. Another way to see this is that in the holographic dual which is the decoupled theory on the brane world-volume, the degrees of freedom (D-strings in type IIB) are stuck near the intersection of the branes and only have kinetic terms in the flat directions. All the states in the theory are therefore singlets under internal rotations in the curved directions. 

Let us fix coordinates so that the flat ones are $x^0...x^{d-1}$. For $n=3$ and $\mu=0$, the above Calabi-Yau is just the ALE space $\IR^4/\IZ_2$. This has an isometry which rotates $(z_2,z_3)$ into each other. Performing a T-duality along this circle gives us two NS5-branes at the origin of $\IR^4$. The deformation of the ALE space into a two-center Taub-Nut is dual to the separation of the location of the 5-branes in the $\IR^4$. This is the $d=6$ theory. We can chose coordinates so that $x^6$ is the circle of isometry. This circle shrinks to zero size at the location of the singularity, and thus in the brane picture, this grows to an infinite direction near the branes. The separation of the branes is in the direction $(x^6,x^8,x^9)$ at $x^7=0$. The former three coordinates can be rotated so that the branes are at $x^{7,8,9}=0$ giving rise to the $SO(3)$ as in the previous section. The exchange of the two 5-branes $x^6 \rightarrow -x^6$ is the symmetry which has a non-trivial $\IZ_2$ action on the bosons. 

For $n=4,5,6$ which corresponds to the lower dimensional theories, we can write the Calabi-Yau near the deformed singularity as $z_1^2 + z_2^2 + z^2 = \mu$ where $z^2 = z_3^2 +.. z_n^2$. This should be thought of as an ALE fibration over the curve $ z^2 = \mu$. We can perform the same T-duality as above, and we get an NS5-brane wrapped on $z^2 = \mu$. For $\mu=0$, the curved part of the configuration is embedded in $\IR^{2(n-2)}$ parameterized by $x^d ... x^5,x^8,x^9$, and the coordinates $x^6,x^7$ are transverse to this brane. When $\mu$ is turned on, there is a deformation in the internal space $\IR^{2(n-2)}$ and in $x^6$ which causes the brane configuration to posses a minimum size $S^{n-3}$. Suppressing all the coordinates except the transverse directions $(x^6,x^7)$, there are two point-like branes at the origin of the plane for $\mu=0$, and the deformation separates them by $\mu$ in this plane. 

Let us now look at the global symmetries. As mentioned above, the only global symmetries are those involving the transverse directions. First let us consider the singular case $\mu=0$. For $d=4,2,0$ ($n=4,5,6$), the symmetry is $U(1) \times U(1)$ of momentum and winding around the circle that rotates the two transverse directions $(x^6,x^7)$. We can identify the rotation of the transverse directions in complex coordinates as $z \rightarrow e^{i \phi} z$, which is produced by $z_i \rightarrow e^{i \phi} z_i$. The deformation preserves the $U(1)$ of winding and breaks the rotation to the discrete rotation $z \rightarrow -z$. It is clear that, upto an $SO(n-2)$ symmetry rotating the branes, this is equivalent to $z_3 \rightarrow -z_3$ for odd $n$ and to a trivial rotation for even $n$. That the former is a rotation by $\pi$ can be seen by looking at the projection of the curve to $(z_4,...,z_n)=0$ which is invariant under the above symmetry. 

From the Calabi-Yau point of view, we can present the following argument. At a particular value of the radius in the geoemtry, the bosonic symmetry group is $O(n) = SO(n) \sdtimes \IZ_2$ for the theory specified by $n$. For odd $n$, the $SO(n) $ commutes with the $\IZ_2$ and the product is really a direct product. In this case, the $\IZ_2$ is just that of parity, and wavefunctions of particles propagating on this manifold are labelled by the quantum numbers of $SO(n)$, and a sign. For even $n$, the $SO(n) $ does not commute with the $\IZ_2$, and the good quantum numbers are those of the $SO(n)$ alone. The wavefunctions with non-zero spin are peaked away from the origin and in the limit of going near the singularity, they decouple. Only the singlets under the $SO(n)$ remain, and only for odd $n$, there is also a $\IZ_2$ symmetry. This $\IZ_2$ symmetry acts on the top form of the Calabi-Yau as $\Omega \rightarrow -\Omega$, which is equivalent to a rotation of $\pi$. These facts can be checked in the cases that there are explicit metrics which have been written down. For instance, in the conifold theory [see e.g. \kleb\ and  refs. therein] , the angle $\psi$ has period $4\pi$ in the full conifold. Near the tip though, a rotation by $2\pi$ along with a rotation in the sphere brings you back to the same point. 

We end with two comments:
\item{1.} The difference between the $d=6$ case and the lower dimensional ones is the following: for $n=3$, turning on  $\mu$ deforms the curve $z_3^2=0$  in the directions $(x^8,x^9)$ and in $x^6$ just like the other cases. Because the curve in this case is a pair of points, the symmetry structure is $SO(4) \rightarrow O(3)$. In the other cases, because the two directions $(x^8,x^9)$ are filled in by the brane, the symmetry is $U(1) \times U(1) \rightarrow U(1) \times \IZ_2$. The deformation in the $d=4,2,0$ theories correspond to turning on the sine-Liouville term in the action. As we saw earlier, the enhanced symmetry in the $d=6$ theory could be seen from the conformal field theory as well where the sine-Liouville and cigar terms are related by the symmetry.
\item{2.} When $n$ is even $(d=4,0)$, we saw that the there is only a rotation by $2 \pi$ in the curved directions of the brane that remains as a symmetry. The fermions pick up a negative sign under this and this gives the discrete winding symmetry which is equivalent to $(-)^{F_S}$ for $d=4$. For $d=0$, the supercharges are charged under this rotation by $2 \pi$, but because of the lack of flat directions, the physical bosons and fermions are not paired by the supercharges. The physical fermionic operators in the lists \iibspectrum\ , \iiaspectrum\ are thus not charged under the discrete winding symmetry.

\vskip 1cm

\centerline{\bf Acknowledgements}
I would like to thank N.~Seiberg for his large contribution to this work, from suggesting the problem to providing constant guidance, many useful comments and support at all stages of the work. I would also like to thank S.~Ashok, C.~Beasley, D.~Berenstein, S.~Cherkis, A.~Hashimoto, J.~Maldacena, L.~Maoz, J.~McGreevy, I.~Klebanov, S.~Mukhi, P.~Ouyang, M.~Rangamani, N.~Saulina and K.~Skenderis for useful and enjoyable discussions. I would like to thank L.~Maoz, J.~McGreevy and M.~Rangamani for useful comments on a preliminary draft. This research was supported in part by the National Science Foundation Grants No. PHY-9802484 and PHY-0243680. Any opinions, findings, and conclusions or recommendations expressed in this material are those of the author and do not necessarily reflect the views of the National Science Foundation.

\appendix{A}{Spectrum of the higher dimensional theories}

The vertex operators become complicated as the mass in spacetime increases. To construct the states at a general level, we can use the oscillator algebra of the worldsheet fields which are free asymptotically. If we fix the reparameterization invariance on the worldsheet by going to the light cone gauge, we can write down the form of the general state in the theory asymptotically in terms of worldsheet oscillators in $d$ transverse directions. 
\centerline{Table 1: {\it Type $0$ spectrum}}
\medskip
\centerline{\vbox{\offinterlineskip
\hrule
\halign{\vrule # & \strut\ \hfil #\ \hfil & \vrule # & \ # \ & \vrule # & \ # \ & \vrule # & \ # \ & \vrule # \cr
height3pt&\omit&&&&&&&\cr 
&{\bf Theory}&&{\bf Sector} &&${\bf (n,w)}$ &&\hfil {\bf Example ($d=4$)}\hfil & \cr
height3pt&\omit&&&&&&&\cr 
\noalign{\hrule}
height3pt&\omit&&&&&&&\cr 
& $0B$ and $0A$ && NS+NS+ && $(N,M)$ &&  Graviton multiplet $(G,B,\phi)$& \cr
&&& NS-NS- && $(N,M)$ && Tachyonic scalar $T$ & \cr 
height2pt&\omit&&&&&&&\cr 
\noalign{\hrule}
height2pt&\omit&&&&&&&\cr 
&$0B$ && R+R+ && $(N,M)$ && Scalar, two form  $(\phi_1,B_{ab}^{sd})$ &\cr 
&&& R-R- & &$(N,M)$ && Scalar, two form  $(\phi_2,B_{ab}^{asd})$ &\cr 
height2pt&\omit&&&&&&&\cr 
\noalign{\hrule}
height2pt&\omit&&&&&&&\cr 
&$0A$&& R+R- && $(N,M)$ && Vector $A^1_a$  &\cr 
&&& R-R+ && $(N,M)$ && Vector $A^2_a$ &\cr 
height3pt&\omit&&&&&&&\cr }
\hrule
}}

\medskip
\leftskip=35pt
\rightskip=35pt
\baselineskip=12pt

\ndt The operators are labelled by the momentum $n$ and winding $w$ ($N,M$ integers). The $\pm$ signs in the sectors are the naive chirality on the worldsheet defined asymptotically. These are not good quantum numbers, and they are only indicated to classify the spin of a particle in the asymptotic region. The superscripts in the R-R sector indicate the (anti)self duality condition that the field strenghts of the listed potential obeys. If there is no superscript, the field strength is unconstrained.

\leftskip=0pt
\rightskip=0pt
\baselineskip=18pt

\bigskip

\centerline{Table 2: {\it Type II spectrum ($d=4$)}}
\medskip
\centerline{\vbox{\offinterlineskip
\hrule
\halign{\vrule # & \strut\ \hfil #\ \hfil & \vrule # & \ # \ & \vrule # & \ # \ & \vrule # & \ # \ & \vrule # \cr
height3pt&\omit&&&&&&&\cr 
&{\bf Theory}&&{\bf Sector} &&${\bf (n,w)}$ &&\hfil {\bf Example ($d=4$)}\hfil & \cr
height3pt&\omit&&&&&&&\cr 
\noalign{\hrule}
height3pt&\omit&&&&&&&\cr 
& II$B$ and II$A$ && NS+NS+ && $(2N,2M),(2N+1,2M+1)$ && $(G,B,\phi)$ & \cr
&&& NS-NS- && $(2N+1,2M),(2N,2M+1)$ && $T$ (non-tachyonic)  & \cr 
height2pt&\omit&&&&&&&\cr 
\noalign{\hrule}
height2pt&\omit&&&&&&&\cr 
&II$B$ && R+R+ && $(2N,2M),(2N+1,2M+1)$ && $(\phi_1,B_{ab}^{sd})$  &\cr 
&&& R-R- & & $(2N+1,2M),(2N,2M+1)$ && $(\phi_2,B_{ab}^{asd})$  &\cr 
height2pt&\omit&&&&&&&\cr 
\noalign{\hrule}
height2pt&\omit&&&&&&&\cr 
&II$A$&& R+R- && $(2N,2M),(2N+1,2M+1)$ && $A^1_a$  &\cr 
&&& R-R+ && $(2N+1,2M),(2N,2M+1)$ && $A^2_a$ &\cr 
height2pt&\omit&&&&&&&\cr 
\noalign{\hrule}
height2pt&\omit&&&&&&&\cr 
&II$B$&& R+NS-, NS+R-  &&  $(2N+\half,2M-\half),(2N-\half,2M+\half)$  && (Supercharge, \hfil &\cr
&&& R-NS+, NS-R+  &&   $(2N+\half,2M+\half),(2N-\half,2M-\half)$  && \hfil  Photino, Gravitino)  &\cr 
height2pt&\omit&&&&&&&\cr 
\noalign{\hrule}
height2pt&\omit&&&&&&&\cr 
&II$A$&& R+NS-, NS+R+  && $(2N+\half,2M-\half),(2N-\half,2M+\half)$    &&  (Supercharge, \hfil &\cr
&&& R-NS+, NS-R-  &&  $(2N+\half,2M+\half),(2N-\half,2M-\half)$    &&  \hfil  Photino, Gravitino)  &\cr 
height3pt&\omit&&&&&&&\cr }
\hrule
}}

\medskip
\leftskip=35pt
\rightskip=35pt
\baselineskip=12pt

\ndt The notation is as in Table 1. Notice that the particles cannot be described by an effective action, because not all the modes are present. The fermionic states are obtained by the action of the supercharges on the bosons. Their Lorentz representation is correllated with the momentum.

\leftskip=0pt
\rightskip=0pt
\baselineskip=18pt

\appendix{B}{Partition functions of the various theories}

On the sphere, the partition function vanishes in all even dimensions due to the presence of two extra fermionic zero modes on the cigar \kutsei. To compute the torus partition sum, we shall go to light cone gauge. In this gauge in $d$ dimensional theories, the physical oscillators are $\alpha_n^I,\t\alpha_n^I, \psi_n^I, \t\psi_n^I$, $I=\rho,\theta,1,2,..d-2$. 

A general state is built out of the raising operators acting on the vacuum, and has momentum $p,k_\mu$ in the $d+1$ non-compact directions and momentum $n$ and winding $w$ around the circle. The left and right moving momentum around the circle are given by $k_L={Q n\over 2}+{w\over Q};\;k_R={Q n\over 2}-{w\over Q}$. The physical state condition gives us the mass of a state in terms of the oscillators. In the left moving sector (the right moving mass shell equation is defined likewise):
\eqn\mshellosc{\eqalign{
&L_0 := \sum_{n=1}^\infty n \alpha^I_{-n}\alpha^I_{n} + \sum_{r=\half}^\infty r \psi^I_{-r}\psi^I_r +\half(k_\mu^2 +k_L^2 -p^2) - {d\over 16} = 0 \qquad {\rm (NS)} \cr
&L_0 := \sum_{n=1}^\infty n \alpha^I_{-n}\alpha^I_{n} + \sum_{m=1}^\infty m \psi^I_{-m}\psi^I_m +\half(k_\mu^2 +k_L^2 -p^2) = 0 \qquad {\rm (R)} \cr
}}
The energy of the NS sector vacuum state (tachyon) can be understood as arising from the presence of the dilaton, or the curvature of spacetime, as in the previous sections.\foot{As a reference, in the 10 dimensional string theory, the energy of the NS vacuum in our units is $-\half$.} It can also be seen by adding the zero point energies of $d$ free bosons and $d$ antiperiodic fermions on the worldsheet. The energy of the R sector vacuum is $0$. 

The one loop partition sum is  split into the integration over the non-compact momenta, the sum over the allowed compact momentum and winding, and oscillator sums; taking into account the GSO projection:
\eqn\defpartfn{
 {\bf Z_{T^2}}=V_{d+2} \int {d^2 \tau \over 4\tau_2} \int {d^d k^\mu \over (2\pi)^d} \int
{dp \over 4\pi} \sum_{n,w} {1 \over R} \Tr (-1)^{F_S} q^{L_0} \qb^{\bar L_0}
}
where $q=e^{2\pi i \tau}$ and the trace is over the physical Hilbert space of the theories.

The definition of the $\Theta$ and $\eta$ functions are:
\eqn\defeta{\eqalign{
 &\eta(\tau)=q^{1\over 24} \prod_{n=1}^\infty (1-q^n) \cr
 &\Theta_{ab}(\tau)=q^{a\over 8} e^{\pi i ab/2} \prod_{n=1}^\infty (1-q^n) (1+e^{\pi i b}q^{n-(1-a)/2})(1+e^{-\pi i b}q^{n-(1+a)/2}) \cr
}}
The modular invariance of the partition sum of the type $0$ theories is straightforward in all cases, relying on the transformation properties of the $\Theta$ functions defined above. 

The partition function for the type $0'$ and type II theories is computed by taking into account the GSO projection. We present below the modular invariant partition sums for the theories as a function of the dimension. 

\subsec{d=0} 
In this case, fixing to light-cone gauge leaves us with no oscillators. As we found in the text, there are only the few field-theoretic degrees of freedom in the pure black hole, and we can compute the partition function purely as a sum over winding and momenta, and integration over the dilaton direction. In the type $0$ theory, we find:
\eqn\pfntwodO{
 {\bf Z_{T^2}} \sim V_2 \int {d^2 \tau \over \tau_2^2}  \sum_{m,w=-\infty}^\infty \exp\left({-\pi |m-w\tau|^2\over 2 \tau_2}\right) 
}
In the type II theory, the sum over the NS-NS sector cancels the sum over the spacetime fermions, and the RR field has all momentum and winding (we do not impose the Dirac equation at this level). The full partition function is therefore half of the type 0 partition function:
\eqn\pfntwod{
 {\bf Z_{T^2}} \sim V_2 \int {d^2 \tau \over \tau_2^2}  \sum_{m,w=-\infty}^\infty \exp\left({-\pi |m-w\tau|^2\over 2 \tau_2}\right) 
}

\subsec{d=2}

\ndt The torus partition function of the type 0 theory is: 
\eqn\fourdpartfn{\eqalign{
 {\bf Z_{T^2}} \sim V_4 \int {d^2 \tau \over \tau_2^2} &{1 \over \tau_2} \sum_{m,w=-\infty}^\infty \exp\left({-2 \pi |m-w\tau|^2\over 3 \tau_2}\right) {1\over |\eta(\tau)|^4} \times \cr 
&\left[ 
\left|{\Theta_{00}(\tau)\over \eta(\tau)}\right|^2 +
\left|{\Theta_{01}(\tau)\over \eta(\tau)}\right|^2 +
\left|{\Theta_{10}(\tau)\over \eta(\tau)}\right|^2 \mp
\left|{\Theta_{11}(\tau)\over \eta(\tau)}\right|^2
\right]\cr
}}
The partition function of the type $0'$ theory is given by:
\eqn\fourdpfnOpr{\eqalign{
 {\bf Z_{T^2}} \sim V_4 \int {d^2 \tau \over \tau_2^2} &{1 \over \tau_2^\half} {1\over |\eta(\tau)|^2} \times \cr 
&\sum_{k=0}^2 \left[ 
\left|{\Theta_{00}(\tau) \Theta_{{2k\over3}0}(3\tau)\over \eta^2(\tau)} \right|^2 - 
\left|{\Theta_{01}(\tau) \Theta_{{2k\over3}1}(3\tau)\over \eta^2(\tau)} \right|^2+
\left|{\Theta_{10}(\tau) \Theta_{{2k+1\over3}0}(3\tau)\over \eta^2(\tau)} \right|^2 
\right]\cr
}}
The modular invariance of the above function can be seen by noticing the identity \farkas\
\eqn\iden{
\Theta_{0,{2k\over 3}}({-1\over 3\tau})=\sum_{l=0}^{l=2} \Theta_{{2l\over 3},2k}({-3\over \tau}).
}
The partition function for the type II theory is given by:
\eqn\fourdIIpfn{
 {\bf Z_{T^2}} \sim V_4 \int {d^2 \tau \over \tau_2^2} {1 \over \tau_2^\half} {1\over |\eta(\tau)|^2} \times 
\left|
{\Theta_{00}(\tau) \Theta_{00}({\tau\over3})\over \eta(\tau)^2} -
{\Theta_{01}(\tau) \Theta_{01}({\tau\over3})\over \eta(\tau)^2} -
{\Theta_{10}(\tau) \Theta_{10}({\tau\over3})\over \eta(\tau)^2}
\right|^2 
}
This function is modular invariant \farkas , and in fact, it vanishes, as it should for a supersymmetric theory. 

\subsec{d=4} 
The type $0$ partition sum is 
\eqn\partfn{\eqalign{
 {\bf Z_{T^2}} \sim V_6 \int {d^2 \tau \over \tau_2^2} &{1 \over \tau_2^2} \sum_{m,w=-\infty}^\infty \exp\left({-\pi |m-w\tau|^2\over \tau_2}\right) {1\over |\eta(\tau)|^8} \times \cr 
&\left[ 
\left|{\Theta_{00}(\tau)\over \eta(\tau)}\right|^4 +
\left|{\Theta_{01}(\tau)\over \eta(\tau)}\right|^4 +
\left|{\Theta_{10}(\tau)\over \eta(\tau)}\right|^4 \mp
\left|{\Theta_{11}(\tau)\over \eta(\tau)}\right|^4 
\right]\cr
}}
where the $\mp$ sign is for the 0A and 0B theories respectively. The tachyon is contained in the sum of the first two terms.

To fill a gap in the text, we will use this example to get an estimate of the high energy behaviour of our theories. This calculation can easily be generalized to the other cases. In the left moving R sector with no momentum, the partition sum is
\eqn\roughsum{
 \sum d_n q^n = \Tr q^{L_0} = 4 \prod_{n=1}^\infty \left({1+q^n \over 1-q^n}\right)^4 {\sim_{q\rightarrow 1}} (1-q)^2 \exp\left({\pi^2 \over 1-q}\right)
}
The chiral density of states $d_n$ for large $n$ can be estimated by a saddle point approximation of the contour integral, and it is given by
$ d^{Op}_n \sim n^{-{7\over 4}} \exp(2\pi\sqrt{n})$. 
In the closed string, we have 
$ d^{Cl}_n = (d^{Op}_n)^2  \sim n^{-{7\over 2}} \exp(4\pi\sqrt{n})$. Considering that the mass of the string is related to the level as ${\alpha'\over 4} m^2=n$ gives the mass density of states:
\eqn\massdos{
 \rho(m) \sim m^{-6}\exp({m\over m_0}), \qquad m_0 = (\pi\sqrt{4\alpha'})
}

The partition sum on the torus (for IIB and IIA) is
\eqn\typetwopfn{\eqalign{
 {\bf Z_{T^2}} \sim V_6 \int {d^2 \tau \over \tau_2^2} &{1 \over \tau_2^{3/2}} {1\over |\eta(\tau)|^6} \times 
\left[ \left| \left({\Theta_{00}^2(\tau) \over \eta^2(\tau)} + {\Theta_{01}^2(\tau)\over \eta^2(\tau)}\right) {\alpha_{11}(\tau) \over \eta(\tau)} - {\Theta_{10}^2(\tau)\over \eta^2(\tau)}  {\alpha_{01}(\tau) \over \eta(\tau)} \right|^2 \right. \cr
& \left. + \left| \left({\Theta_{00}^2(\tau)\over \eta^2(\tau)} -{\Theta_{01}^2(\tau)\over \eta^2(\tau)} \right) {\alpha_{01}(\tau) \over \eta(\tau)} - {\Theta_{10}^2(\tau)\over \eta^2(\tau)} {\alpha_{11}(\tau) \over \eta(\tau)} \right|^2 \right] \cr
}}
where the functions $\alpha_{m1}$ are:
\eqn\defalpha{
 \alpha_{m1}(\tau)= \sum_{n \in {\bf Z}} q^{(n+m/2)^2} \;\; (m=0,1).
}
The partition function defined above vanishes as expected for spacetime supersymmetric theories \mizo. The first expression in the oscillator sum contains the tachyon (with odd momenta) and the tachyno, and the second contains the graviton multiplet (with even momenta) and its superpartner.

\subsec{d=6}
The torus partition function of the type 0 theory is: 
\eqn\eightdpartfn{\eqalign{
 {\bf Z_{T^2}} \sim V_8 \int {d^2 \tau \over \tau_2^2} &{1 \over \tau_2^3} \sum_{m,w=-\infty}^\infty \exp\left({-2 \pi |m-w\tau|^2\over \tau_2}\right) {1\over |\eta(\tau)|^{12}} \times \cr 
&\left[ 
\left|{\Theta_{00}(\tau)\over \eta(\tau)}\right|^6 +
\left|{\Theta_{01}(\tau)\over \eta(\tau)}\right|^6 +
\left|{\Theta_{10}(\tau)\over \eta(\tau)}\right|^6 \mp
\left|{\Theta_{11}(\tau)\over \eta(\tau)}\right|^6
\right]\cr
}}
The partition function of the type $0'$ theory (which is the type $0$ theory on two NS5-branes is given by:
\eqn\eightdpfnOpr{
 {\bf Z_{T^2}} \sim V_8 \int {d^2 \tau \over \tau_2^2} {1 \over \tau_2^{5\over 2}} {1\over |\eta(\tau)|^{10}} \times 
\left[ 
\left|{\Theta_{00}(\tau)\over \eta(\tau)} \right|^8+
\left|{\Theta_{01}(\tau)\over \eta(\tau)} \right|^8+
\left|{\Theta_{10}(\tau)\over \eta(\tau)} \right|^8 
\right]
} 
The partition function for the type II theories is given by:
\eqn\fourdIIpfn{
 {\bf Z_{T^2}} \sim V_8 \int {d^2 \tau \over \tau_2^2} {1 \over \tau_2^{5\over 2}} {1\over |\eta(\tau)|^{10}} \times 
\left|
{\Theta_{00}^4(\tau) \over \eta^4(\tau)} -
{\Theta_{01}^4(\tau) \over \eta^4(\tau)} -
{\Theta_{10}^4(\tau) \over \eta^4(\tau)}
\right|^2 
}

\appendix{C}{Quantization in the Green-Schwarz formalism ($d=4$)}

The six bosonic coordinates are $X^a = X^a + \t X^a$ and their spacetime superpartners are $\theta_\alpha$, $\t \theta_\alpha$ which are six dimensional Weyl spinors in the ${\bf 4}$ or ${\bf {\bar 4}}$ of $SU(4)$. We shall work in light cone gauge where the coordinates are $X^I$ ($I=\rho,\theta,1,2$), and Weyl spinors $S_\alpha,{\bar S_\alpha}$ in the ${\bf 2}$ and $\t S_\alpha, {\bar {\t S_\alpha}}$ in the ${\bf 2}$ or ${\bf 2'}$ (II$B$ or II$A$) of $SO(4)$. In the T-dual variables, the interaction is the sine-Liouville interaction:
\eqn\gsaction{\eqalign{
&S= \int d^2\sigma \; \left(\half \p_\mu X^I \p^\mu X^I - i {\bar S^\alpha} \p_+ S^\alpha -i {\bar {\t S^\alpha}} \p_- \t S^\alpha \right. \cr
&  \left. \qquad \qquad - e^{-\irt2 (\rho+ i\theta)} 
S^2 {\t S}^2 - e^{-\irt2 (\rho- i\theta)} {\bar S^2} {\bar {\t S}}^2
+ e^{-\rt2 \rho}  \right).
}}
In the above action, a specific choice of $\sigma$ matrices has been made which breaks the $U(1)$ symmetry which rotates the spinors $S^1, S^2$ into each other. 

The right and left moving Lorentz generators are given by:
\eqn\lorentz{\eqalign{
& J^{12}= X^1 P^2 - P^2 X^1 + \irt2 {\dot \theta} - \half {\bar S^2} S^2 + \half {\bar S^1} S^1, \cr
& \t J^{12}= \t X^1 \t P^2 - \t P^2 \t X^1 + \irt2 {\dot {\t \theta}} - \half {\bar {\t S^2}} \t S^2 + \half {\bar {\t S^1}} \t S^1, \cr
}}
The rotation in the 1-2 direction is given the sum of the left and right moving parts which also we shall call $J^{12}$. The only other physical symmetry of the theory (consistent with the R-NS analysis) is:
\eqn\conscurr{\eqalign{
& w = \irt2 ({\dot \theta} - \dot {\t \theta}) - \half ( {\bar S^2} S^2 - {\bar {\t S^2}} \t S^2) - \half ( {\bar S^1} S^1 - {\bar {\t S^1}} \t S^1) \cr 
}}
In the cigar theory, this current is the conserved momentum which we saw earlier. Note that in this formalism, the above conserved current is a combination of naive momemtum around the cigar and a piece which acts on the worldsheet fermions.

The free action has eight leftmoving (and another eight rightmoving) supersymmetries - four shifts of the spinors and four linearly realised supersymmetries. The interaction preserves only half of the above symmetries - only shifts of $S^1, {\bar S^1}$ are preserved, and two of the linear supersymmetries (which are modified due to the superpotential so that they are no longer purely left or right moving) are preserved. These supercharges are charged under the currents $J^{12}, w$ (here $\sigma^3 = \left(\matrix{-1 &0 \cr 0 &1}\right)$):
\eqn\jscomm{\eqalign{
& [J^{12},Q^\alpha] = \half \sigma^3_{\alpha\beta} Q^\beta,\qquad  [J^{12}, {\t Q^\alpha}] = \half \sigma^3_{\alpha\beta} {\t Q^\beta} ;\cr
& [J^{12},\bar Q^\alpha] = -\half \sigma^3_{\alpha\beta} \bar Q^\beta,\qquad  [J^{12}, \bar {\t Q^\alpha}] = -\half \sigma^3_{\alpha\beta} \bar {\t Q^\beta} ;\cr
& [w,Q^\alpha] = \half Q^\alpha,\qquad  [w,{\t Q^\alpha}] = -\half {\t Q^\alpha} ;\cr
& [w,\bar Q^\alpha] = -\half \bar Q^\alpha,\qquad  [w,\bar {\t Q^\alpha}] = \half \bar {\t Q^\alpha}. \cr
}}

The ground state is annihilated by all bosonic and fermionic lowering operators, and is in a irrep of the susy algebra which (for the left moving part alone) consists of four states, all of vanishing energy. Below, we list the states and the charges under the two conserved currents winding and the spacetime rotation $J^{12}$. We start with a state which asymptotically obeys
${\bar S^\alpha}|0\rangle = {\dot \theta} |0\rangle = 0$, and build three other left moving states of the same energy (and repeat the construction on the rightmoving side):

\centerline{Table 5: Ground states in the Green-Schwarz formalism}
\bigskip
\centerline{\vbox{\offinterlineskip
\hrule
\halign{\vrule # & \strut\ \hfil $#$\ \hfil & \vrule # & \ $#$ \ &  \ $#$ \ & \ $#$ \ & \ $#$ \ & \vrule # & \ $#$ \ &  \ $#$ \ & \ $#$ \ & \ $#$ \ & \vrule # \cr
height3pt&\omit&&&&&&&&&&& \cr
&{\bf{\rm State}} && |0\rangle & {\bar S^2}|0\rangle &{\bar S^1}|0\rangle &{\bar S^2}{\bar S^1}|0\rangle && |0\rangle  &\bar {\t S^2}|0\rangle &\bar {\t S^1}|0\rangle &\bar {\t S^2} \bar {\t S^1}|0\rangle & \cr
height2pt&\omit&&&&&&&&&&& \cr
\noalign{\hrule}
height2pt&\omit&&&&&&&&&&& \cr
&{\bf J^{12}} &&0 & -\half & \half & 0 && 0 & -\half & \half & 0 &\cr
&{\bf w} &&0 & -\half & -\half & -1 && 0 & \half & \half & 1 &\cr
height3pt&\omit&&&&&&&&&&&\cr}
\hrule
}}
\medskip

The closed string spectrum is built by tensoring these two together. The spacetime spin properties are given by the charge under $J^{12}$. At this level, we get 16 states with zero energy which form a scalar(tachyon) multiplet. The four tachyon states have winding number $-1,0,0,1$. The conserved charges are the same as in the NSR formalism.

The states with least non-zero energy $E=\half$ is built on states which obeys ${\bar S^\alpha}|\pm1\rangle=0,\; {\dot \theta} |\pm1\rangle=\pm \irt2 |\pm1\rangle$ tensored with the rightmovers. This gives $64$ states with spins corresponding to the graviton multiplet. The gravitons have zero winding, and the conserved charges of all the states are the same as in the NSR formalism.  

The general state in the spectrum is constructed by the action of the raising operators of the four bosons $\alpha^I_{-n}, {\t \alpha}^I_{-n}$ and the eight fermions $S^{\alpha}_{-n}, {\bar S^{\alpha}_n}, {\t S}^{\alpha}_{-n}, {\bar {\t S}^{\alpha}_n}$, on the states annihilated by all the lowering operators with arbitrary integer or half integer value of momemtum. The algebra of the conserved charges being the same as in the NSR formalism, and the ground state being the same guarantees that the full spectrum is identical as well.

\appendix{D}{Conformal invariance at second order in the $d=6$ theory}

The operators $\CL^{\mu\nu}$ are all of dimension $(1,1)$. The nine operators generate other operators when they come close to each other, and the coefficients of these operators should vanish for the theory to remain conformal. 

\item{1.} As in the Liouville theory, we assume that in the OPE of $e^{p_1 \rho}$ with $e^{p_2 \rho}$, the operator that is generated is normalizable ($p<0$) if $e^{(p_1+p_2) \rho}$ is. 
\item{2.} Asymptotically, we can consistently choose to keep operators which decay slower than a certain rate (i.e., $p>p_0$ for some $p_0$). In the five brane theory, we can think of this as an expansion in $x^2/r^2$ where $x$ is the separation of the branes and $r$ is the radial distance from the center of mass. 

It can be checked to order $e^{-\rho}$, that the are the only operators that preserve $\CN =(2,2)$ superconformal symmetry on the worldsheet and are not total derivatives are the above nine. This means that the above set of operators are closed under the OPE (upto perhaps change in the zero modes of fields). 

We have then, $A_i (z) A_j(0) \sim {c_1 \over 2z} \epsilon_{ijk} A_k(0)$. The free field part of
this OPE can be explicitly computed, and the dilaton part - $e^{-\rho}$ with itself - gives again
$e^{-\rho}$ with a coefficient $c_1$ which we {\it assume} is non-zero. This relation is
consistent
with charge conservation after taking into account the background charge of the dilaton.

If we consider separating the branes in the directions $(X^8,X^9)$ alone, we turn on only $\CL^{int} = C_{66}\CL^{66} +C_{77}\CL^{77} + 2 C_{67}\CL^{67}$. At second order in the coefficients, there are new operators generated - the contribution to the beta function by these new operators is proportional to $((C_{66}+C_{77}) \CL^{int} + (C_{66}C_{77} - C_{67}^2)\CL^{newint})e^{-\rho-\t \rho}$. Demanding that the second term vanishes gives us $C_{66}C_{77} - C_{67}^2=0$. We get five other such conditions for other separations. 

The solution to these constraints is labelled by four parameters $x^\mu$ such that $C_{\mu\nu}=x_\mu x_\nu$. It can be checked that this is a good solution for the most general interaction allowed. The interaction $\CL^{int} = x_\mu x_\nu \CL^{\mu\nu} e^{-\rho-\t \rho}$ generates at second order a contribution to the beta function proprtional to $(x_\mu x^\mu) \CL^{int}$ which is equivalent to a change in the zero mode of the dilaton. More precisely, the expectation value of $e^{-\rho_0}$ is identified with $(x_\mu x^\mu)$. The value of the dilaton at the tip corresponds to the distance of the branes from the origin, and the other three operators correspond to moving the branes on a sphere at constant radius. 

\listrefs

\end